# SOUND DESIGN STRATEGIES FOR LATENT AUDIO SPACE EXPLORATIONS USING DEEP LEARNING ARCHITECTURES

**Kıvanç Tatar**
Chalmers University of Technology
Sweden
`tatar@chalmers.se`

**Kelsey Cotton**
Chalmers University of Technology
Sweden
`kelsey@chalmers.se`

**Daniel Bisig**
Zurich University of the Arts
Switzerland
`daniel.bisig@zhdk.ch`

## ABSTRACT

The research in Deep Learning applications in sound and music computing have gathered an interest in the recent years; however, there is still a missing link between these new technologies and on how they can be incorporated into real-world artistic practices. In this work, we explore a well-known Deep Learning architecture called Variational Autoencoders (VAEs). These architectures have been used in many areas for generating latent spaces where data points are organized so that similar data points locate closer to each other. Previously, VAEs have been used for generating latent timbre spaces or latent spaces of symbolic music excepts. Applying VAE to audio features of timbre requires a vocoder to transform the timbre generated by the network to an audio signal, which is computationally expensive. In this work, we apply VAEs to raw audio data directly while bypassing audio feature extraction. This approach allows the practitioners to use any audio recording while giving flexibility and control over the aesthetics through dataset curation. The lower computation time in audio signal generation allows the raw audio approach to be incorporated into real-time applications. In this work, we propose three strategies to explore latent spaces of audio and timbre for sound design applications. By doing so, our aim is to initiate a conversation on artistic approaches and strategies to utilize latent audio spaces in sound and music practices.

## 1. INTRODUCTION

A variety of musical practices devote significant theory into the examination of approaches to organizing sound. Such organization has been focused on different musical properties in two main tracks: 1- encompassing rhythmic, melodic, harmonic, and timbral organization in conventional European musical theory, 2- sound organization in a 20th century electro-acoustic music theoretical perspective. The difference in approaches to musical organization assists practitioners in creating strategic, systematic, or theoretical approaches that are assistive to their respective artistic practices, whilst additionally enabling spectators to appreciate or investigate the musical works in detail.

In this paper, we address Machine Learning (ML) and Artificial Intelligence (AI) tools that are geared towards a specific type of organization, based on similarity. Architectures, such as Autoencoders in Deep Learning and UMAP, t-SNE, or k-means clustering in Machine Learning, calculate the distances between data points in a dataset to create an abstract space within which similar data points are organized. Such abstract spaces are often referred to as 'latent spaces' in ML literature. This work focuses on a specific type of latent spaces–latent audio spaces–while investigating the incorporation of latent audio spaces into an artistic practice.

Our paper further seeks to initiate discussion into how latent audio spaces can be explored in musically meaningful ways. Meaning-making of, and with, technology has been a core field of exploration within the third wave of HCI [1–3]. We see this as a similarly meaningful area of inquiry within musical-AI practice. In musical applications, however, the quality dimensions that is meaningful to musical practices are not so clear. As music is embedded in culture and cultural practices, it can be argued that quality dimensions are substantially different across different cultures, societal demographics and geographies. Design utilization of latent spaces in creative exploration of quality dimensions affords an open space for sonic experimentation and discovery of sounds that are particular to the user's own aesthetic and practice (as supplied through their self-provided dataset of sounds).

Our approach in this regard is to create strategies that use audio recordings as a tool to navigate latent audio spaces. The use of algorithmic approaches provide relatively simple means for exploring latent audio spaces between two audio recordings. Starting from such musically curious questions, we utilise algorithmic approaches to explore latent audio spaces generated by Deep Learning architectures, specifically Variational Autoencoders. We propose three latent audio space interpolation strategies [1]. These three interpolation strategies have been developed during an artistic residency of the first author [2], wherein the live-coding practice was utilized to explore latent audio spaces.

Live coding enables unique interaction possibilities for exploring latent spaces using mathematical functions and analytic approaches. In live-coding, the performer codes musical layers and sonic gestures whilst working with relatively longer durations.The real-time planning of composi-tional layers has further connections to comprovisation [4]. Accordingly, the three strategies that we propose in this paper aim to generate long duration compositional layers using a dataset of relatively short audio samples.This paper therefore contributes with: 1) initiating a discussion on how to utilize latent audio spaces in an artistic practices, 2) Proposing three algorithmic approaches for traversing a latent audio space and 3) an open-source latent-timbre synthesis model for implementation within an experimental electronic music practice.

---
[1] The open-source repository of our work can be found at `https://github.com/ktatar/rawaudiovae`
[2] `https://kivanctatar.com/Coding-the-Latent`

## 2. RELATED WORK

This paper is situated within a nexus of similar works which utilise a variety of deep learning architectures.Of specific interest to this paper is existing work encompassing timbral approaches; audio generation with WaveNet and variational autoencoders (VAEs); and integrations of differentiable digital signal processing (DDSP) with deep learning methods.

### 2.1 Timbre based approaches

Previous work has examined implementations of Convolutional Neural Networks (CNN) combined with WaveRNN vocoder to facilitate real-time synthesis of audio using neural networks [5]; the application of a chroma vector to augment input and influence latent space navigation [6]; timbral transfer using Constant Q-Transform (CQT) [7]; and latent timbre synthesis using variational autoencoders [8].

Hantrakul et al [5] have explored varying conditioning features as a means for improving computational speeds of performing synthesis of instrument sounds. They present a combination of CNN timbre synthesis with WaveRNN-based autoregressive vocoder, where the pipeline is conditioned with pitch and amplitude control. Other work from Colonel et al [6] has deployed a Variational Autoencoder for timbral synthesis and interpolation, utilising a one-hot encoded chroma vector to guide the generation with pitch input. Kumar et al's introduction of a generative adversarial network (GAN) architecture for conditional audio synthesis [9] present MelGAN architecture, which is a non-autoregressive convolutional architecture which performs waveform generation, and a novel example of a GAN trained for raw audio generation from MFCC spectrograms, devoid of supplementary distillation and perceptual loss functions. Additional timbral approaches from Huang et al [7] explore timbral transfer between instruments whilst retaining musical content such as pitch, loudness and rhythm. Their presented pipeline TimbreTron performs timbral transfer using a wavelet based timbre feature called Constant Q Transform (CQT). Tatar et al. [8] present a complimentary timbral approach, in their construction of an audio synthesis framework vis-a-vis their Latent Timbre Synthesis (LTS) framework. The core structural components of the LTS framework utilise CQT calculation, deep learning to generate a latent audio space, and inverse synthesis via magnitude CQT spectrograms and Grifin-Lim phase estimation. As revealed through their evaluation, this architecture afforded novel opportunities for interpolation and extrapolation of audio excerpts. Esling et al [10] similarly contribute to this research domain with their construction of a latent space for the analysis and synthesis of audio content whilst further establishing perceptual timbral similarities–compiled via dissimilarity ratings and further analysed with Multi-Dimensional Scaling (MDS)–of a cross section of instruments. This was achieved through the regularisation of variational auto-encoders (VAEs). They introduce their method for descriptor-based synthesis, affording the synthesis of audio material correlating with a descriptor shape whilst additionally preserving timbral structure. This work stands out with its comparison of different timbral features such as MFCCs, CQTs, and Non-Stationary Gabor Transform as features to train VAEs.

### 2.2 Audio generation, WaveNet, and beyond

The seminal work of van den Oord et al. [11] presents WaveNet, an autoregressive deep neural network utilised for raw audio generation. WaveNet's structure builds upon the PixelCNN [12, 13] architecture, utilising a stack of causal convolutional layers. Using dilations of the convolutional layers, WaveNet limits the model's dependencies of future timesteps–which affords modelling of long-range temporal audio dependencies– whilst preserving the resolution of audio input. Of special significance is WaveNet's fidelity, which outperforms existing text-to-speech (TTS) systems and yielded promising results in audio modelling and speech recognition. Additional contributions from Dieleman et al examine [14] the generation of raw audio utilising autoregressive autoencoders (ADAs), comparing their presented argmax autoencoder (AMAE) against vector quantisation variational autoencoders (VQ-VAE) in capturing long-range correlations in waveforms during the generation of audio waveforms. The main drawback of WaveNet and its follow-up architectures is their relatively high computational complexity, which is a barrier to incorporate these architectures into real-world artistic practices.

Approaching audio generation from a multimodal domain-transference perspective, Bisig and Tatar [15] transpose dance pose sequences into audio waveforms, utilising a deep learning architecture which pairs a sequence-to-sequence model and an adversarial autoencoder for raw audio generation from audio encodings. Raw Music from Free Movements (RAMFEM) learns and recreates existing music-movement relationship in the raw audio domain. The pipeline combines two autoencoders, an RNN based sequence-to-sequence network for generating latent human body pose spaces and a VAE for creating latent raw audio spaces. The architecture that we present here builds on the Adverserial VAE in RAMFEM. Although the audio VAE in RAMFEM is using a discriminator network within and Adversarial VAE pipeline, the architecture that we present in this paper omits the discriminator network for a more simplified VAE pipeline.

### 2.3 Differentiable DSP

Offering a contrasting approach, Engel at al [16] have explored digital signal processing integrated within a deep learning framework, utilising a module system of differ-entiable synthesis blocks and vocoders to generate high-fidelity audio. They demonstrate the modularity of models utilising DDSP components in enabling independent control over additional audio features (such as pitch and loudness; modelling and transference of room acoustic); timbral transference between dissimilar instruments; and extrapolation to novel pitch material during synthesis.

## 3. TWO TYPES OF LATENT AUDIO SPACES

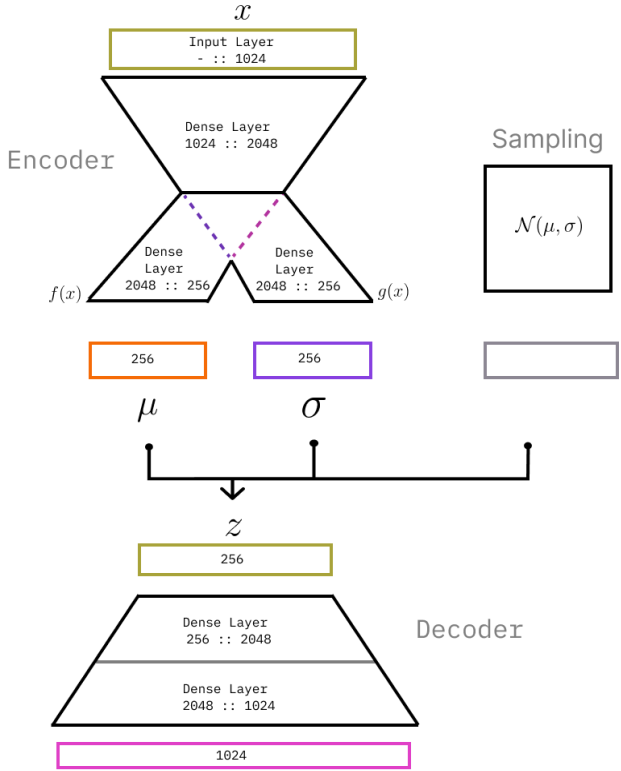

Figure 1: The architecture of *RawAudioVAE*

### 3.1 Continuous Latent Audio Spaces

Continuous latent audio spaces encode audio quanta into a latent space using various Machine Learning approaches, such as Variational Autoencoders (VAEs). In these types of latent audio spaces, the input is an audio window, containing a couple of thousands of samples, and in the durations of fractions of a second. The continuous latent audio spaces encode sonic gestures or audio samples as a continuous path inside the latent space, where each point is encoded from one audio window. In Figure 2, the red circles represent audio windows with 1024 samples that are encoded from a single audio sample file, and the path-like appearance of these circles are an emerging property of the latent audio space approach using VAEs. Hence, the red path is a continuous latent audio space encoding of a sonic gesture.

Autoencoders (AE) are neural networks that learn to reconstruct original data instances by compressing and reconstructing said data into, and from, lower dimensional representations [17]. An AE consists of two networks- an encoder that translates high dimensional input data into a latent encoding, and a decoder that reconstructs the original data from the latent encoding as accurately as possible. To accomplish this task, the autoencoder learns an efficient compression of the data into a latent encoding.

In a vanilla AE, the encoder directly converts a data instance into a deterministic encoding. This approach suffers from two main shortcomings: the latent space produced by the encoder is sparsely populated and contains many regions from which encodings cannot be decoded into meaningful data instances. The similarity metrics of the data space is not preserved in latent space, i.e. encodings of similar data instances may not be situated close to each other within latent space. These two shortcomings make it difficult to search for encodings that can then be decoded into valid data instances in vanilla AEs.

Variational autoencoders (VAE) alleviate these shortcoming by making the latent space less sparse and by imposing similarity metrics that resemble those of the original data space [18]. Unlike AEs, the encoder in a VAE maps a data instance into parameters for a probability distribution, from which a latent code can be sampled. The encoder approximates $p(z|x)$, whereas the decoder's function is to maximize the probability of the posterior $p(x|z)$, where $x$ is the input data and $z$ is a latent vector that represents that input data. VAEs assumes that $z$ is a probability distribution, and the Normal distribution $\mathcal{N}(\mu, \sigma)$ is a common distribution function to sample $z$, where $\mu$ is the mean and $\sigma$ is the standard variance. However, the sampling function is not differentiable. VAE employ a reparameterization trick to circumvent the problem that gradients cannot backpropagate through the stochastic sampling from a probability distribution. This trick expresses the sampling of a random variable as a combination of a deterministic variable and a factor epsilon, which is sampled from a standard normal distribution. Specifically, the encoder outputs two vectors of $\mu$ and $\sigma$ which are passed to the distribution function. The encoder generates latent space by modelling $p(z|x)$ using $q^*(z|x) = \mathcal{N}(z; f(x), g(x)^2 I)$ where $\mu_M = f(x)$, $f \in F$, $\sigma_M = g(x)$, $g \in G$, and $M$ is the number dimensions in the latent space. The decoder's input, $z$ is sampled from $q(z) = \mathcal{N}(z; f(x), g(x)^2 I)$. The loss function in VAE's training includes two terms, the expectation maximization and the regularization term of Kullback-Leibler divergence between $q^*(z|x)$ and $p^*(z)$,

$$L_{f,g} = \mathbb{E}_{q^*(z)}[\log p^*(x|z)] - \alpha \cdot D_{KL}[q^*(z|x)||p^*(z)] \quad (1)$$

VAE are trained by maximizing the probability of generating valid data instances, while enforcing the posterior $q^*(z|x)$ to be close to the prior $p^*(z)$. The similarity between these two distributions is commonly quantified by using the Kullback-Leibler divergence, as in the equation 1. The full pipeline of RawAudioVAE is depicted in Figure 1. The specific VAE model covered in Section 5, is trained for 500 epochs with learning rate of $10^{e-4}$, where $\alpha$ is $10^{e-4}$.

### 3.2 Discrete Latent Audio Spaces

In comparison to continuous latent audio spaces, discrete latent spaces organize sonic gestures that are in the range of a couple of seconds duration. The audio samples in the training dataset are often thumbnailed using a set of audio features. In Music Information Retrieval, a known approach in audio thumbnailing is the bag-of-frames approach, where the mean and standard deviations of a number of audio features are calculated over an audio excerpt or sample. The statistic outcome of these audio feature are brought together as one vector to represent the full audio sample. ML approaches that create discrete latent audio spaces use audio thumbnail vectors and the distances between these vectors to organize the audio samples in a latent space. In cases such as audiostellar [20], each dot in the latent space represent one audio sample. In other ML approaches such as Self-Organizing Maps [21], each square represents a cluster of audio samples (see Figure 3)

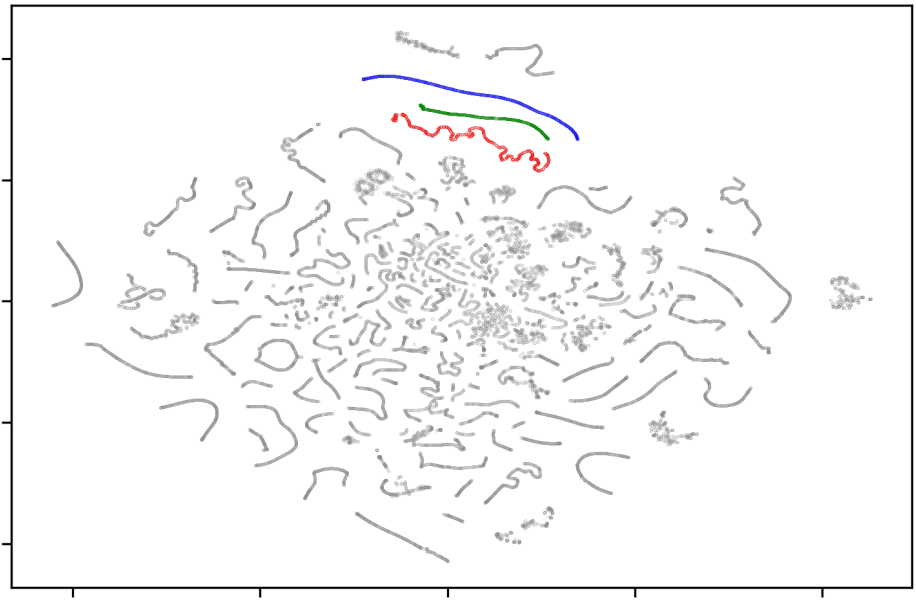

Figure 2: This visualization illustrates an example continuous latent audio space in our previous work Latent Timbre Synthesis [8], where each point is an encoding of one CQT vector. The image is generated by using t-distributed stochastic neighbor embedding to project the latent vectors with into a two-dimensional visualization. The red and blue dots are audio windows from two audio files in the training dataset, whereas the green dots are the mid-way interpolation between the latent vectors of those two audio files

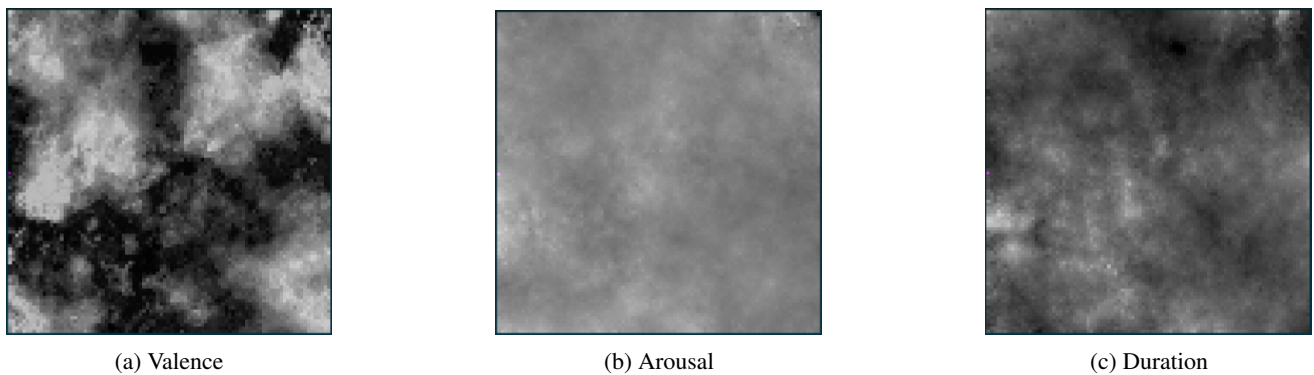

(a) Valence  (b) Arousal  (c) Duration

Figure 3: These three dimensions are from a Self-Organizing Map with 35 dimensions, trained on 53318 audio samples, with 94 by 94 dimensions, used in the first author's previous artwork [19]. The SOM map is utilized in algorithm 3 to generate a longer duration audio with similarity.

## 4. STRATEGIES FOR EXPLORING LATENT SPACES OF AUDIO

In this work, we provide three strategies to explore latent audio spaces. The question as to how to utilize such parametric spaces for audio synthesis is still to be investigated. Currently, knowledge on properties of latent spaces and how to analyze latent audio spaces with considerations of musical qualities and sonicaesthetics is still missing. Mathematical methods to analyze such latent spaces are beneficial to learn about the architecture, hyper-parameter optimization, and helpful to notice the issues in model training, however, they do not necessarily provide guidance as to the aesthetic possibilities of latent audio spaces. There is still a knowledge gap in how to reveal sonic properties and aesthetics of latent audio spaces. Such knowledge could help artists and practitioners to utilize a relatively large parametric space in real-world artistic applications.

Furthermore, we lack clear methods for investigating the sonic aesthetics of latent sound spaces, and have an absence of articulated musical quality [3] taxonomies that would help to identify aesthetic dimensions for investigation in a latent sound space. Instead of focusing on approaches where VAE is conditioned with additional dimensions, as per the work by Colonel and Keene [6], our approach instead focuses on explorations of latent sound spaces between two selected sounds. Guiding the latent audio space exploration with audio inputs allow the flexibility to explore different aesthetics in a variety of musical contexts and cultures, whilst avoiding the clarification of a musical dimension for conditioning.

The three strategies below utilize two audio input files, in which the user explores the latent spaces in-between those two files. The audio inputs are set to have the same duration. The audio inputs constrain the aesthetics of the latent space exploration. In all methods, the input files are encoded using the VAE encoder to generate their latent vectors. The audio files are windowed first, and the VAE input is a single audio window of 1024 samples. Each audio window is encoded separately. This non-autoregressive approach decreases the computational complexity significantly, and simultaneously enables real-time implementation. The interpolations are carried out by generating latent vectors between the latent vector pairs of audio input files (see Figure 4 and 5).

### 4.1 Stepwise Interpolations

**Algorithm 1** Stepwise Interpolations

1: **Load** a trained network pair of an $Encoder$ and a $Decoder$
2: **Load** two audio files, $input_1$ and $input_2$
3: **if** $size(input_1) < size(input_2)$ **then**
4:   $input_2 = input_2[0 : size(input_1)]$
5: **else if** **then**
6:   $input_1 = input_1[0 : size(input_2)]$
7: **end if**
8: $l\_mean_{input_1}, l\_std_{input_1} = Encoder(input_1)$
9: $l\_mean_{input_2}, l\_std_{input_2} = Encoder(input_2)$
10: **Define** a range, $r \in \mathbb{R}$ and a step size, $s \in \mathbb{R}$
11: **for all** $i \leftarrow 0$ to $\frac{r}{s}$ **do**
12:   $l\_mean_{a_i} = (i*s)*l\_mean_{input_1} + (1-(i*s))*l\_mean_{input_2}$
13:   $l\_std_{a_i} = (i*s)*l\_std_{input_1} + (1-(i*s))*l\_std_{input_2}$
14:   $p = \mathcal{N}(l\_mean_{a_i}, l\_std_{a_i})$
15:   $a_i = Decoder(p)$
16: **end for**
17: $a_{out} = [a_0, a_1, \ldots, a_{(\frac{r}{s}+1)}]$

The first method utilizes a fixed interpolation amount across the timeline. That is, the multiplier $s$ on line 12 and 13 in algorithm 1 is a constant, not a vector. In this approach, we utilize short audio samples with a duration of a couple of seconds. The longer audio file out of the two chosen audio files, is first shortened so that both files are the same duration. It is required to have the same number of audio windows in both audio files. VAE represent each audio window as a normal distribution $\mathcal{N}(\mu, \sigma)$ in the latent space, where $\mu$ is the mean and the $\sigma$ is the standard deviation. Each audio window is passed through the VAE encoder to generate the latent vectors. The encoder outputs two vectors per audio window, a vector for $\mu$, and another for $\sigma$. Each vector is the size of the latent space dimensions- which is 256 dimensions in our case. The mean and standard deviation pairs are passed to a normal distribution function (line 14 in algorithm 1), which is then passed to the decoder. The usage of normal distribution is referred as the reparameterization trick, which allow us to synthesize new sounds by exploring latent vectors in between two latent vectors of audio files from the training dataset.

---

[3] Our understanding of "qualities" here is founded in how this term is used within the studies and methodologies of the third wave of HCI.

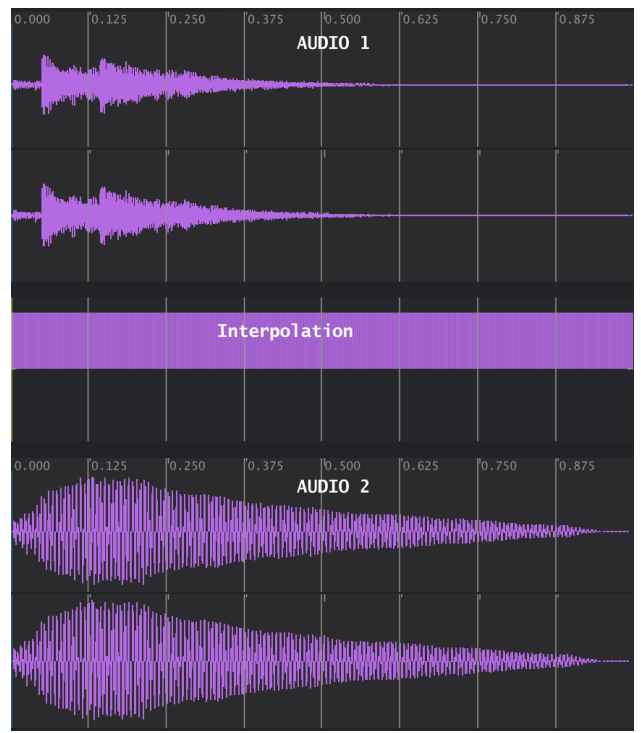

Figure 4: Illustration of a static interpolation amount to explore latent spaces between two audio files

In the second half of the algorithm 1, we apply a simple grid search of interpolation amounts. For each interpolation amount, the new latent vectors are generated using the formula on line 12 and 13. For example, the interpolation amount 0.2 means that 80% of latent vectors from audio 1 is added to the 20% of latent vectors from audio 2. Essentially, if we would define a multidimensional plane in between two latent vectors, the interpolation amount 0.2 would refer to a region that is four times more distant to one sound then to the other. By focusing on a range a static interpolation amounts, we can start from a reconstruction of one audio input file, gradually changing to the other audio file by changing the interpolation amount at each step. Hence, this strategy allows the user to listen to the latent space and how the sound shifts in the latent space of a trained *RawAudioVAE* model.

### 4.2 Interpolations in Meso-scale

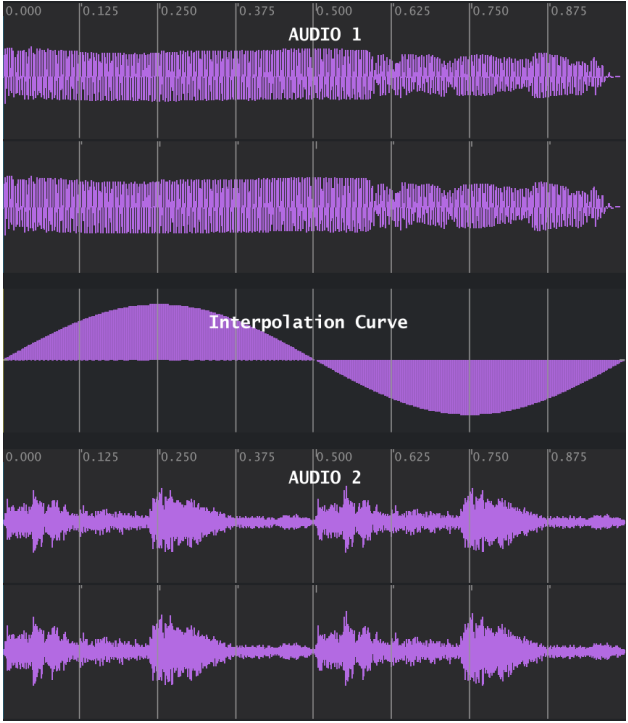

Figure 5: Illustration of a dynamic interpolation curve to explore latent spaces between two audio files

The second strategy in the exploration of latent audio space differs from our first in two regards. The first difference is in how audio input files are created with an additional step, *prior* to the latent space exploration. In this second strategy, we create two relatively long audio excerpts by concatenating similar audio samples in the training dataset. The audio similarity is carried out by utilizing audio clustering, vis-a-vis a Self-Organizing Map trained on a thumbnail of audio samples in the training dataset. Further details of this clustering process can be found in our previous work [21, 22]. To create two audio excerpts with a relatively long duration, we choose two audio clusters from the Self-Organizing Map, concatenating all files in those clusters to create one single audio file with relatively long duration. The clustering helps to constrain the sonic aesthetics in the audio file, and improves the consistency in the concatenated audio file, compared to a random selection of audio samples for the concatenation.

Generating two input audio files through concatenation of audio samples enables us to work with audio files of 10-30 seconds in duration. Audio files within such duration are beneficial to this second interpolation strategy, because the interpolation amount is dynamically changing as the timeline progresses. In our previous work, we carried out a study where composer participants designed dynamic interpolation amounts to explore a latent audio space for sonic explorations in their compositional practice [8]. A qualitative finding from this earlier study was that com-

**Algorithm 2** Interpolations in Meso-scale

1: **Load** a trained network pair of an $Encoder$ and a $Decoder$
2: **Select** two SOM clusters, $c_1$ and $c_2$
3: **Define** $input_1 = []$ and $input_2 = []$
4: **for all** $a_{sample} \in c_1$ **do**
5: $\quad input_1 = [input_1, a_{sample}]$
6: **end for**
7: **for all** $a_{sample} \in c_2$ **do**
8: $\quad input_2 = [input_2, a_{sample}]$
9: **end for**
10: **if** $size(input_1) < size(input_2)$ **then**
11: $\quad input_2 = input_2[0 : size(input_1)]$
12: **else if** **then**
13: $\quad input_1 = input_1[0 : size(input_2)]$
14: **end if**
15: $l\_mean_{input_1}, l\_std_{input_1} = Encoder(input_1)$
16: $l\_mean_{input_2}, l\_std_{input_2} = Encoder(input_2)$
17: **Generate** a signal, $curve$ **where** $max(curve) \leq 1$ and $min(curve) \geq -1$ and $size(curve) = size(input_1)$
18: $l\_mean_{a_{new}} = curve * l\_mean_{input_1} + (1 - curve) * l\_mean_{input_2}$
19: $l\_std_{a_{new}} = curve * l\_std_{input_1} + (1 - curve) * l\_std_{input_2}$
20: $p = \mathcal{N}(l\_mean_{a_{new}}, l\_std_{a_{new}})$
21: $a_{out} = Decoder(p)$

posers found the design of interpolation curves to explore latent audio spaces between two input audio files to be musically relevant.

The notion of dynamic interpolation in this second strategy differs from our first strategy (where we had used a static interpolation amount). The dynamic interpolation approach of our second strategy applies different interpolation amounts to each audio window pairs, wherein one audio window is taken from each input audio file. Across the audio timeline, the interpolation amount follows a curve the size of the total number of audio windows in the given audio input file. A variety of mathematical functions can be applied here to generate interpolation curves, such as sinusoidal functions with different periods etc.[4]

### 4.3 Interpolations with Extensions

Similar to our second interpolation strategy, the third strategy utilizes a cluster of audio samples to concatenate audio samples within the same cluster to generate a longer duration audio file with sonic consistency. However, our third strategy *further* increases the duration of the concatenated audio using a windowing trick. This last strategy emerged from an error in using the *AudioDataset* class instead of *TestDataset* class when loading audio input files to be fed to the VAE. The difference between those two classes is in how they treat the audio windowing given an audio file in the training dataset. By using a 1024-sample windowing with a hop size of 256 samples, we can stretch the audio

---
[4] Please refer to the section *Interpolations in Meso-scale* in the supplementary jupyter notebook code.

**Algorithm 3** Interpolations with extensions

1: **Load** a trained network pair of an $Encoder$ and a $Decoder$
2: **Select** two SOM clusters, $c_1$ and $c_2$
3: **Define** $input_1 = []$ and $input_2 = []$
4: **for all** $a_{sample} \in c_1$ **do**
5:     $input_1 = [input_1, a_{sample}]$
6: **end for**
7: **for all** $a_{sample} \in c_2$ **do**
8:     $input_2 = [input_2, a_{sample}]$
9: **end for**
10: **if** $size(input_1) < size(input_2)$ **then**
11:     $input_2 = input_2[0 : size(input_1)]$
12: **else if** **then**
13:     $input_1 = input_1[0 : size(input_2)]$
14: **end if**
15: **Define** $input_{1s} = []$ and $input_{2s} = []$
16: **for all** $i \in \mathbb{R} \leftarrow 0$ to $\frac{size(input_1)}{hop\_size}$) **do**
17:     $segment = input_1[i, i + window\_size]$
18:     $input_{1s} = [input_{1s}, segment]$
19: **end for**
20: **for all** $i \in \mathbb{R} \leftarrow 0$ to $\frac{size(input_2)}{hop\_size}$) **do**
21:     $segment = input_2[i, i + window\_size]$
22:     $input_{2s} = [input_{2s}, segment]$
23: **end for**
24: $l\_mean_{input_1s}, l\_std_{input_1s} = Encoder(input_1s)$
25: $l\_mean_{input_2s}, l\_std_{input_2s} = Encoder(input_2s)$
26: **Generate** a signal, $curve$ where $max(curve) \leq 1$ and $min(curve) \geq -1$ and $size(curve) = size(input_1s)$
27: $l\_mean_{a_i} = curve * l\_mean_{input_1s} + (1 - curve) * l\_mean_{input_2s}$
28: $l\_std_{a_i} = curve * l\_std_{input_1s} + (1 - curve) * l\_std_{input_2s}$
29: $p = \mathcal{N}(l\_mean_{a_i}, l\_std_{a_i})$
30: $a_{out} = Decoder(p)$

file to four times its duration (see algortihm 3 line 16–23). As in the second strategy, this third strategy applies a dynamic interpolation approach, where an interpolation curve applies different interpolation amounts to different sections of the audio timeline.

## 5. COMPROVISATION IN LIVE CODING USING INTERPOLATION STRATEGIES IN THE LATENT AUDIO SPACE

The three strategies that we mention in Section 4 emerged during the first author's artist residency at the Center for Art and Media (ZKM) Karlsruhe [5] . The residency topic was exploring artistic ways of utilizing latent audio spaces generated by Deep Learning architectures in live coding performance. Live coding provides a unique way to interact with latent audio spaces, as it is possible to access a trained Deep Learning model and explore the latent space of such a model using math functions and analytical ap-proaches. This approach differs substantially in comparison to the direct interaction and manipulation approaches in performances with Digital Musical Instruments (DMIs). The nature of live coding shifts the performance mindset from action-perception loops to a comprovisation [4], wherein the live coder handles multiple musical layers at once whilst planning, coding, synthesizing or generating future musical actions and layers. The interaction between the performer and their tools in this particular application of live coding is unique, since the performance mindset shifts the focus towards long-term planning of musical composition. Put simply, it takes too long to change the code synthesizing sounds from the latent space of a Deep Learning model in Python jupyter notebooks. To overcome this temporal issue, we looked into strategies in which the performer could synthesize long durational musical layers with ease and speed. With this concern, all of the strategies that we presented in this paper aimed to synthesize long duration audio files, even though the training dataset of the DL model consisted of short duration audio samples (sometimes only a few seconds long). This technological constraint–quick real-time code manipulations to synthesize sound from the latent space of the DL model–afforded an interesting live coding performance mindset where the performer–the first author–has created a musical vocabulary and strategies for comprovisation within the live performance.

---

[5] More info and the live performance can be found at `https://kivanctatar.com/Coding-the-Latent`

## 6. CONCLUSIONS AND FUTURE WORK

This work has emerged from our musically curious question towards Machine Learning algorithms generating latent spaces and how these algorithms can be applied to raw audio data. We clarified two types of latent audio spaces–continuous and discrete–that appear in the literature, and which were utilized in this work. We proposed three simple algorithmic strategies that can be applied to explore latent audio spaces within artistic practices such as live coding performances. In our future work, we aim to compare a variety of latent audio spaces including timbral and raw audio approaches, with respect to their mathematical properties. Looking into analytical methods in ML to analyze latent audio spaces can inform possibilities of sonic aesthetics and musical qualities that could be enabled by incorporating latent audio approaches in to musical practices. The evaluation of these frameworks within music creation and performance contexts is also a priority. Lastly, the framework that we use in this work, *RawAudioVAE* is lightweight, and can generate a 1-second audio in less than 10ms. This low computational complexity makes *RawAudioVAE* a promising architecture to be interfaced with other technologies within the designs of Digital Musical Instruments in future.

### Acknowledgments

This work was partially supported by the Wallenberg AI, Autonomous Systems and Software Program – Humanities and Society (WASP-HS) funded by the Marianne and Marcus Wallenberg Foundation and the Marcus and Amalia Wallenberg Foundation. Additionally, this research was previously supported by the Swiss National Science Foundation, and Canada Council for the Arts.